\documentclass{article}
\pdfoutput=1

\usepackage[final, nonatbib]{neurips_2024}

\usepackage[utf8]{inputenc} %
\usepackage[T1]{fontenc}    %
\usepackage{hyperref}       %
\usepackage{url}            %
\usepackage{booktabs}       %
\usepackage{nicefrac}       %
\usepackage{microtype}      %
\usepackage{graphicx}
\usepackage{pdfpages}

\usepackage{thmtools}
\usepackage{thm-restate}

\usepackage{tikz}
\usepackage{pgf}
\usetikzlibrary{patterns}
\usepackage{environ}
\usepackage{float}

\usepackage{caption}

\usepackage{multirow}
\usepackage{xcolor}
\usepackage{xspace}
\usepackage{dirtytalk}
\usepackage{amsmath}
\usepackage{amssymb}
\usepackage{amsthm}

\usepackage{hyperref}
\usepackage{url}
\usepackage{microtype}
\usepackage{relsize}

\usepackage{algorithm}
\usepackage{algorithmicx}
\usepackage{algpseudocode}
\usepackage{arydshln}

\usepackage{listings}
\usepackage{xspace}
\usepackage{multirow}

\algtext*{EndWhile}%
\algtext*{EndIf}%
\algtext*{EndFor}%

\newcommand{\parameters}[1]{\item[\textbf{Parameters:}] \textit{#1}}
\newcommand\nonumberinc{\refstepcounter{equation}\nonumber}

\usepackage{colortbl}

\usepackage{amsmath}
\usepackage{amsfonts}       %
\usepackage{amssymb}
\usepackage{amsthm}
\usepackage{mathtools}
\usepackage{nicefrac}
\usepackage[capitalize,noabbrev,nameinlink]{cleveref} %

\usepackage{xcolor, soul}
    \definecolor{orangeish}{HTML}{f1a340} %
    \definecolor{grayish}{HTML}{f7f7f7} %
    \definecolor{purpleish}{HTML}{998ec3} %
    \definecolor{blueish}{HTML}{004488}
    \definecolor{darkgreen}{RGB}{2,100,64} %
    \definecolor{lightgreen}{HTML}{b2f2bb}
    \definecolor{lightblueish}{RGB}{230,244,255}
\PassOptionsToPackage{table,svgnames,dvipsnames}{xcolor}

\usepackage{hyperref} %

\hypersetup{
colorlinks=true,
linkcolor=darkgray,
citecolor=gray,
pdftitle={Faster Private Minimum Spanning Trees},
pdfsubject={Differential Privacy, Spanning Trees},
pdfauthor={Rasmus Pagh, Lukas Retschmeier},
pdfkeywords={Differential Privacy, Graphs, Data Structures}
}

\theoremstyle{plain}
\newtheorem{theorem}{Theorem}[section]

\newtheorem{definition}[theorem]{Definition}
\newtheorem{lemma}[theorem]{Lemma}
\newtheorem{corollary}[theorem]{Corollary}
\theoremstyle{definition}

\theoremstyle{remark}

\usepackage[textsize=tiny]{todonotes}
\setuptodonotes{color=blue!25,linecolor=blue}

\title{Faster Private Minimum Spanning Trees}

\author{%
  Rasmus Pagh\\
  BARC, University of Copenhagen\\
  Universitetsparken 1, Copenhagen \\
  \texttt{pagh@di.ku.dk} \\
  \And
  Lukas  Retschmeier \\
  BARC, University of Copenhagen \\
  Universitetsparken 1, Copenhagen\\
  \texttt{lure@di.ku.dk} \\
}

\renewcommand{\epsilon}{\varepsilon}

\setul{0.5ex}{0.3ex}
\newcommand{\ulcolor}[2][Red]{\setulcolor{#1}\ul{#2}}

\DeclareCaptionFormat{custom}
{%
    \textbf{#1#2}\textit{\small #3}
}
\Crefformat{figure}{#2Fig.~#1#3}
\Crefmultiformat{figure}{Figs.~#2#1#3}{ and~#2#1#3}{, #2#1#3}{ and~#2#1#3}

\newcommand{\rnmCompleteFast}{\textsc{Faster-Discretized-Report-Noisy-Max}\xspace}

\newcommand{\zcdp}{zCDP\xspace}

\newcommand{\R}{\ensuremath{\mathbb{R}}}
\newcommand{\E}{\ensuremath{E}} %

\newcommand{\G}{\ensuremath{G=(V,\E)}\xspace}

\newcommand{\Prob}{\ensuremath{\mathbb{P}}} %

\newcommand{\expD}{\ensuremath{{\tt Exp}}}
\newcommand{\maxExpD}{\ensuremath{\tt MaxExp}}

\newcommand{\UniformD}{\ensuremath{\tt Uni}} %
\newcommand{\Uni}{\ensuremath{{\tt Uni}}} %
\newcommand{\lapD}{\ensuremath{{\tt Lap}}}
\newcommand{\binomialD}{\ensuremath{{\tt Bin}}}
\newcommand{\normalExpD}{\ensuremath{{{\tt N}}}}

\newcommand{\round}[2]{{#1}\left\lfloor \dfrac{{#2}}{#1} \right\rfloor}
\newcommand{\roundInline}[2]{{#1}\left\lfloor \frac{{#2}}{#1} \right\rfloor}

\newcommand{\errorPrivatePrimS}{\ensuremath{\mathcal{O}\left(n^{\nicefrac{3}{2}}\log (n) \Delta_\infty / \sqrt{\rho}\right)}\xspace}
\newcommand{\errorPrivatePrimSD}{\ensuremath{\mathcal{O}\left(n^{\nicefrac{3}{2}}\log (n)/\sqrt{\rho}\right)}\xspace}

\newcommand{\errorGaussianNoise}{\ensuremath{\mathcal{O}\left(n^2 \sqrt{\log (n)/\rho}\right)}\xspace}

\newcommand{\errorGaussianNoiselone}{\ensuremath{\mathcal{O}\left(n \sqrt{\log (n)/\rho}\right)}\xspace}
\newcommand{\errorGaussianNoiseloneExact}{\ensuremath{\mathcal{O}\left(n \sqrt{\log (n/\gamma)/\rho}\right)}\xspace}

\newcommand{\errorGaussianNoiselinftyExact}{\ensuremath{\mathcal{O}\left(n^2 \sqrt{\log (n/\gamma)/\rho}\right)}\xspace}

\crefname{lstlisting}{Listing}{Listings}
\newcommand{\basicCodeStyle}{\ttfamily\small}
\newcommand{\keywordCodeStyle}{\bfseries}
\newcommand{\builtInCodeStyle}{\itshape\color{orangeish}}
\newcommand{\typesCodeStyle}{\ttfamily\small\color{blueish}}

\lstdefinelanguage{plan}{
	classoffset=0,
	morekeywords={True, False, return, null, if, in, while, do, else, case, break, continue, for, def, log, sqrt, norm},
		keywordstyle=\keywordCodeStyle,
	classoffset=1,
	morekeywords={privQuant, recenter,  scale, clip, noise},
	keywordstyle=\typesCodeStyle,
	classoffset=2,
	morekeywords={estimateStd, rankError, divideBudget, size, scope, center},
	keywordstyle=\builtInCodeStyle,
	classoffset=0,
	morecomment=[s]{/*}{*/},
	morecomment=[l]//,
	morestring=[b]",
	morestring=[b]'
}

\renewcommand\vec{\mathbf}

\DeclareMathOperator*{\argmax}{\tt arg\,max}

\begin{document}

\maketitle

\begin{abstract}

Motivated by applications in clustering and synthetic data generation, we consider the problem of releasing a minimum spanning tree (MST) under \emph{edge-weight} differential privacy constraints where a graph topology $\G$ with $n$ vertices and $m$ edges is public,  the weight matrix $\vec{W}\in \R^{n \times n}$ is private, and we wish to release an approximate MST under $\rho$-zero-concentrated differential privacy.
Weight matrices are considered neighboring if they differ by at most $\Delta_\infty$ in each entry, i.e., we consider an $\ell_\infty$ neighboring relationship.
Existing private MST algorithms either add noise to each entry in $\vec{W}$ and estimate the MST by \emph{post-processing} or add noise to weights \emph{in-place} during the execution of a specific MST algorithm.
Using the post-processing approach with an efficient MST algorithm takes $\mathcal{O}(n^2)$ time on dense graphs but results in an additive error on the weight of the MST of magnitude $\mathcal{O}(n^2\log n)$.
In-place algorithms give asymptotically better utility, but the running time of existing in-place algorithms is $\mathcal{O}(n^3)$ for dense graphs. %
Our main result is a new differentially private MST algorithm that matches the utility of existing in-place methods while running in time $\mathcal{O}(n^2)$ and more generally time $\mathcal{O}(m + n^{3/2}\log n)$ for fixed privacy parameter~$\rho$.
The technical core of our algorithm is an efficient sublinear time simulation of {\tt Report-Noisy-Max} that works by discretizing all edge weights to a multiple of $\Delta_\infty$ and forming groups of edges with identical weights.
Specifically, we present a data structure that allows us to sample a noisy minimum weight edge among at most $\mathcal{O}(n^2)$ cut edges in $\mathcal{O}(\sqrt{n}\log n)$ time. %
Experimental evaluations support our claims that our algorithm significantly improves previous algorithms either in utility or in running time.
\looseness=-1
\end{abstract}

\section{Introduction}

The \emph{minimum spanning tree} (MST) problem is a classical optimization problem with many applications.
In machine learning it is used, among other things, as a clustering algorithm~\cite{DBLP:journals/ida/LaiRN09,DBLP:conf/nips/BateniBDHKLM17,Pinot_2018,jayaram2024massively} and as a subroutine for computing graphical models such as Chow-Liu trees~\cite{chow_liu_1068,McKenna_Miklau_Sheldon_2021}.

We consider the problem of releasing an MST for a graph $\G$ with a public edge set and a private weight matrix $\vec{W}\subseteq \R^{n\times n}$, where $V=\{1,\dots,n\}$.
We consider the $\ell_\infty$ neighborhood relation on the set of weight matrices, where a pair of matrices are considered neighboring if the weight of every edge differs by at most the sensitivity parameter $\Delta_\infty$.
For example, the underlying graph might model a city's metro network where the edge weights represent passenger data, where any individual could have used many different connections.
In this case, we don't want to hide the network; rather, we want to protect the private information encoded in the weights.
A different example comes from graphical modeling: 
Consider a dataset $D = (\mathbf{x}^{(1)},...,\mathbf{x}^{(n)})$ of size $n$ where each vector $\mathbf{x}^{(i)}\in\{0,1\}^d$ represents a list of sensitive binary attributes. 
The Chow-Liu tree is the minimum spanning tree of the negated \emph{mutual information matrix} that encodes the mutual information between every pair of attributes.
Changing one vector $\vec{x}$ in $D$ could simultaneously alter \textit{all} weights by the sensitivity of mutual information, $\Delta_\infty \approx \ln(n)/n$.

Though our techniques apply in general we focus primarily on \emph{dense} graphs where $|E| = \Omega(|V|^2)$.
We want to privately release the edges of a spanning tree $T$ that minimizes the difference between the sum of the edge weights in $T$ and the weight of a minimum spanning tree $T^*$.

\subsection{Previous Work} 

\begin{table*}[t]
    \centering
    \resizebox{0.99\columnwidth}{!}{%
    \begin{tabular}{cc|llcc}
        \textbf{Neighborhood}                 & \textbf{Privacy}                 & \textbf{Reference} & \textbf{Category}     & \textbf{Error} & \textbf{Running time} \\ \hline
        \multirow{3}{*}{$\ell_1$}     & \multirow{2}{*}{$\epsilon$-DP}            & Sealfon \cite{Sealfon_2016}            & Post-processing         & $\mathcal{O}(n \log (n) / \epsilon)$                 &      $\mathcal{O}(n^2)$                  \\
                                  &                                  & Hladík and Tětek \cite{hladik_tetek_2024} & Lower bound &     $\Omega(n \log (n) / \epsilon)$                                 &       ---            \\ \cdashline{3-6} 
                                  & \multirow{1}{*}{$\rho$-zCDP} &       Sealfon \cite{Sealfon_2016}            &  Post-processing       &  \errorGaussianNoiselone           & $\mathcal{O}(n^2)$                         \\ \cline{1-6}
         \multirow{7}{*}{$\ell_\infty$}& \multirow{3}{*}{$\epsilon$-DP} & Sealfon \cite{Sealfon_2016} & Post-processing                                    & $\mathcal{O}(n^3 \log (n) / \epsilon)$     &    $\mathcal{O}(n^2)$               \\
         &           & Pinot (PAMST) \cite{Pinot_2018}            &    In-place                                     & $\mathcal{O}(n^2 \log (n) / \epsilon)$     &    $\mathcal{O}(n^3)$               \\
                                  &                                  & Hladík and Tětek \cite{hladik_tetek_2024} & Lower bound &  $\Omega(n^2 \log (n) /\epsilon)$                                 &  ---                   \\[0.3em] \cdashline{3-6} 
                                  & \multirow{3}{*}{$\rho$-zCDP}          &      Sealfon \cite{Sealfon_2016}          & Post-processing &     \errorGaussianNoise             &     $\mathcal{O}(n^2)$                \\[0.3em]
                                  &          &     Pinot (PAMST) \cite{Pinot_2018}          & In-Place &     \errorPrivatePrimSD             &     $\mathcal{O}(n^3)$                \\[0.3em]
                                  &                                  & \cellcolor{lightblueish}\textbf{New Result} (Fast-PAMST)                                        & \cellcolor{lightblueish}In-place &     \cellcolor{lightblueish}\errorPrivatePrimSD             & \cellcolor{lightblueish}$\mathcal{O}(n^2/\sqrt{\rho})$          \\[0.3em]
                                  &                                  & Wu \cite{wu_2024}                                        & Lower bound&     $\Omega(n^{3/2}\log(n) /\sqrt{\rho})$             &  ---          \\[0.3em] \cline{1-6} 
        \end{tabular}
    }\\
    \caption[short]{Existing bounds for releasing a private MST for an $n$-vertex graph with private weights %
    under $\ell_1$ and $\ell_\infty$ neighboring relations. 
    Error bounds are differences between the weight of the reported spanning tree (with respect to the original weights) and the weight of a minimum spanning tree.
    For simplicity, we list the error for the sensitivity 1 case where neighboring weight matrices can differ by at most 1 in $\ell_1$/$\ell_\infty$ norm ---
    In the general case, all upper and lower bounds scale linearly and with sensitivity.
    Though Sealfon~\cite{Sealfon_2016} considers only $\ell_1$ neighborhoods and $\varepsilon$-DP, the results can be extended easily to $\rho$-zCDP and $\ell_\infty$ neighborhoods.
    Similarly, Pinot~\cite{Pinot_2018} considers only pure differential privacy, but it is simple to do an analysis for \zcdp.
    The time and error bounds of our new algorithm hold with high probability.
    }\label{tab:summ}
\end{table*}

We focus on results for private MST under $\ell_1$ and $\ell_\infty$ neighborhood notions.
An overview of existing and new results can be found in \cref{tab:summ}.
Pinot \cite{Pinot_2018} places existing algorithms into two categories:

\begin{itemize}
    \item \textit{Post-processing} algorithms release a private weight matrix and compute an MST using the noisy weights.
Privacy is ensured by the fundamental property of post-processing \cite{Dwork_Nissim_Smith_2006,bun_steinke_2016}.
Sealfon \cite{Sealfon_2016} was the first to analyze post-processing algorithms, adding Laplace noise to $\vec{W}$ to achieve $\varepsilon$-differential privacy.
The running time of this approach is dominated by the $\mathcal{O}(n^2)$ time it takes to add noise to each entry in $\vec{W}$, since the MST of any graph can be computed in time $\mathcal{O}(n^2)$.
Although they run in linear time for dense graphs, the drawback of these approaches is that the magnitude of the noise is large, $\mathcal{O}(n^2\log n)$.
\item \textit{In-place} algorithms are more sophisticated but use significantly less noise than post-processing algorithms.
The idea is to inject noise whenever a weight is accessed during the execution of a concrete MST algorithm. 
Known representatives are based on Prim-Jarník's algorithm \cite{Prim1957,Pinot_2018}, and Kruskal's algorithm \cite{kruskal_1956, mitrovic_bun_krause_karbasi_2017}.
Both techniques start with an empty set of edges and iteratively grow a set of edges guaranteed to be a subset of an MST.
Therefore, they greedily select the lightest new edge between cuts that respect the edges already chosen in each step.
This can be achieved by using any private selection mechanism, for instance, Report-Noisy-Max~\cite{dwork_roth_2014}, Permute-and-Flip~\cite{mckenna_sheldon_2020}, or the Exponential Mechanism~\cite{dwork_roth_2014}.
Currently, state-of-the-art {\it in-place} algorithms are inefficient because, in each step, they have to consider all edges in the graph to add noise. 
The privacy argument follows from the composition theorems of differential privacy.
\end{itemize}

\subsection{Our Contributions}

We present a new algorithm for the private MST problem under edge-weight differential privacy and the $\ell_\infty$ neighboring relationship that runs in time ${\mathcal{O}(m + n^{3/2}\log n / \sqrt{\rho})}$ and releases a private MST with an error of at most $\errorPrivatePrimS$ with high probability.
For dense graphs and $\rho = \Omega (\log^2(n)/n)$, this implies $O(n^2)$ running time, which is linear in the size of the input.
Our algorithm can be seen as an efficient implementation of PAMST \cite{pinot_2018_ma}.

Our main technique is an efficient simulation of {\tt Report-Noisy-Max} together with a special data structure that can be used inside the Prim-Jarník algorithm to sample the noisy minimum edge in $\mathcal{O}\left(\sqrt{n} \log n / \sqrt{\rho}\right)$ time with high probability, while still being able to insert/delete edges into/from the cut in constant time per edge.
More specifically, we will prove the following:

\begin{samepage}
\begin{restatable}{theorem}{main}
    \label{thm:main}
   There exists a $\mathcal{O}(m + n^{3/2}\log(n) /\sqrt{\rho})$-time $\rho$-zCDP algorithm that, given a public graph topology $G=(V, E)$ with $n$ vertices and $m$ edges together with private weights $\vec{W}$ and sensitivity $\Delta_\infty$, releases the edges of a spanning tree whose weight differs from the weight of the minimum spanning by at most $\errorPrivatePrimS$ with high probability.
\end{restatable}
\end{samepage}

It is worth mentioning that PAMST and our algorithm are both asymptotically optimal for the $\ell_\infty$ neighboring relationship and under $\rho$-zCDP due to a recently found lower bound by Wu \cite{wu_2024}. 

\paragraph{Technical Overview} 

Starting with an empty edge set, the Prim-Jarník algorithm \cite{jarnik1931,Prim1957} iteratively grows a tree that is guaranteed to be a subset of an MST by greedily selecting the lightest possible new edge to add in each step.
It considers the edges between the current tree and the rest of the graph and adds the minimum weight edge in this cut.
To achieve privacy, one can use any private selection mechanism to add an edge having \emph{approximately} the lightest weight in each step \cite{Pinot_2018,mckenna_sheldon_2020}, for instance, Report-Noisy-Max \cite{dwork_roth_2014}, Permute-and-Flip \cite{mckenna_sheldon_2020}, or the Exponential Mechanism \cite{dwork_roth_2014}.
All vanilla versions instantiate noise for each possible output, which generally increases the runtime to $\mathcal{O}(n^3)$ if the graph is dense.

To address the issue, we introduce an efficient simulation technique of {\tt Report-Noisy-Max (RNM)} with exponential noise. 
Recall that {\tt RNM} adds noise to each element and then releases the $\argmax$ among them \cite{dwork_roth_2014}.
It can easily be seen that to find a \textit{minimum} weight edge, we can report a noisy \textit{maximum} by negating all weights first.
We will show how to simulate it efficiently by first rounding all the edge weights to an integer multiple of $\Delta_\infty$ and then putting those with the same value together into groups $F_i$.
We show that we only need to sample one noise term per group because it is enough to sample the maximum noise $Z_i \sim \maxExpD(|F_i|, \lambda)$ for each  $F_i$.
Furthermore, by discretizing to an integer multiple of $\Delta_\infty$, we restrict the number of groups we need to sample $Z_i$ for: By a concentration bound, instead of considering all at most $n^{2}$ groups, it is enough to sample the noise for the topmost $c \sqrt{n} \log n$ groups for some constant $c>0$.
To preserve some probability that edges with weight further away than $M = \sqrt{n} \log n \Delta_\infty$ from the maximum can be sampled, we show that we can treat all those edges as a single, large group $L$, where the number of noise terms $k$ that exceed $M$ is distributed according to $\binomialD(|L|, \exp(-\lambda M))$, because each edge gets noise drawn from $\expD(\lambda)$.
Therefore, we first sample $k$, then generate $k$ noise terms from $\expD(\lambda)_{|\geq M}$, and finally to add them to a uniformly drawn random subset in $L$.
A visualization can be seen in \cref{fig:ouralg}.

\paragraph{Limitations and Broader Impact}
Our main motivation lies in settings where the $\ell_\infty$ neighboring relation is natural (e.g., corresponds to changing one data point in an underlying dataset).
In other settings, it may be more natural to consider $\ell_1$ or $\ell_2$ neighboring relation, and our algorithm may add much more noise than necessary.
Our work is of theoretical nature and we are not aware of any ethical or negative societal consequences.

\section{Preliminaries}\label{sec:background}
We state the most important definitions here and refer for the rest to \cref{apx:defs}.
We always consider a public (connected) graph topology $G = (V,E)$ where $V = \{1, \cdots, n\}$,  $E$ the set of undirected edges and a private weight matrix $\vec{W}\subseteq \R^{n\times n}$.
We often represent $\vec{W}$ also as some function $w \in \R^E$ and denote the cost of a subset $E' \subseteq E$ of edges as $W_{E'} := \sum_{e\in E'}w_e$.

Extending the work of Dwork, McSherry, and Nissim \cite{Dwork_Nissim_Smith_2006}, we will use the (edge-weight) differential privacy framework introduced by Sealfon \cite{Sealfon_2016} and state our results in the $\rho$-zCDP framework proposed by Bun and  Steinke \cite{bun_steinke_2016}.
In this particular setting, a graph topology $G$ is public, and we want to keep the weights $\vec{W}$ private.
Other common notions of privacy for graphs are \emph{edge-level} \cite{hay_li_miklau_jensen_2009} privacy and \emph{node-level} \cite{Kasiviswanathan_nissim_sofya_smith_2013} privacy.

\begin{definition}[$l_p$-neighboring graphs]
   For any graph topology and two weight matrices $\vec{W}, \vec{W'} \in \R^{n \times n}$, we say that they are $l_p$-neighboring (denoted as $W \sim_p W'$), if for any fixed constant $C \in \R: || \vec{W} - \vec{W'} ||_p := \sqrt[\leftroot{-2}\uproot{2}p]{\sum_{i, j\in [n]}\left|w_{i,j}- w_{i,j}' \right|^p} \leq C$.
    In the case of $l_\infty$, we define $|| \vec{W} - \vec{W'} ||_\infty {:= \max_{i, j \in [n]} \left|w_{i,j}-w_{i,j}'\right|} \leq C = \Delta_\infty$
    Hence, we say that two graphs $G = (V, E, \vec{W})$ and $G' = (V, E, \vec{W'})$ are $\ell_p$-neighboring (also denoted $G \sim G'$), if $\vec{W} \sim_p \vec{W'}$.
\end{definition}

\textbf{Sampling Max of Exponentials}
We define the distribution of the maximum $Y$ of $k$ independently drawn exponential random variables $Y = \max\limits_{X_1,\cdots,X_k\sim \expD(\lambda)} \{X_i\}$ as $\maxExpD(k, \lambda)$.

\begin{lemma}\label{def:maxexpd}
For any scaling parameter $\lambda$, the CDF of $\maxExpD(k, \lambda)$, the maximum of $k$ iid exponentially distributed random variables drawn from the same distribution $\expD(\lambda)$ is $F(x; \lambda; k) = (1-e^{x/\lambda})^k)$ for $x \geq 0$ and $0$ otherwise.
\end{lemma}

\begin{proof}
${{\forall Z\in \R}: {{\Pr\limits_{X_i\sim \expD(\lambda)}} \left[\max(X_1, \cdots, X_k) \leq Z\right] } = 
    \left(\Pr\limits_{X\sim \expD(\lambda)}[X \leq Z]\right)^k = 
    {{(1-e^{Z/\lambda})^k}}\qedhere}$ 
 \end{proof}
We can directly draw from that distribution ${X \sim \UniformD(0,1)}$ and transforming it to ${Y = -\lambda \ln(1-\sqrt[k]{X})}$, which works because ${\Pr\left[X \leq t\right] = {\Pr[-\lambda \ln(1-\sqrt[k]{Y})\leq t]} = {\Pr[Y \leq (1-e^{t/\lambda})^{k}]} ={(1-e^{t/\lambda})^{k}}}$.

\paragraph{Report-Noisy-Max (RNM)\label{def:RNM} } 
Given a finite set of candidates $E'$, a dataset $\vec{W}$, and an utility function $w: E \rightarrow \R$ for each $r \in E'$, the \emph{differential private selection problem} asks for an approximately largest item. \cite{dwork_roth_2014} {\tt RNM} is a standard approach, where, for a given dataset $\vec{W}$, one releases a noisy maximum $M = \argmax_{e \in E} w(e, \vec{W}) + \expD(\frac{\epsilon}{2\Delta_\infty})$ with respect to a given function ${w:E \rightarrow \R^d}$ with sensitivity $\Delta_\infty$ (See \cref{psensitivefunction}).
It can be shown that $\Pr\left[\argmax_{e\in E}w(e,\vec{W}) - M \geq \frac{\ln(|E|+t)}{\lambda}\right] \leq e^{-t}$ with probability $t \geq 0$.\label{rnm:utility}
The \emph{Permute-and-Flip} mechanism has been shown to be equal to {\tt RNM}, and {\tt RNM} with Gumbel noise matches exactly the \emph{Exponential Mechanism} \cite{mckenna_sheldon_2020, Ding_Kifer_Sayed_Steinke_Wang_Xiao_Zhang_2021}.

\subsection{Releasing a Minimum Spanning Tree under Edge-DP}

We quickly summarize existing results (see also \cref{tab:summ}), and analyze the Gaussian mechanism using zCDP. 
Corresponding proofs can be found in \cref{ch:analysis}.

First shown by Sealfon \cite{Sealfon_2016}, adding Laplace noise $\lapD(\Delta_1/\epsilon)$ to the weights $\vec{W}\in \R^{n \times n}$ is $\epsilon$-DP and gives an error of at most $\mathcal{O}(n \log (n) /\epsilon)$.
Hladík and Tětek \cite{hladik_tetek_2024} raised the lower bound for pure DP to $\Theta(n \log(n))$, and therefore, Sealfon's mechanism is optimal for the $\ell_1$ neighboring relationship.
In the $\ell_\infty$ case, we have to add noise from $\lapD(n^{2}/\epsilon)$, which makes it impractical as the error now grows cubic in $n$.
If we relax the privacy constraints from pure to approximate DP, 
Although not explicitly mentioned, the following is folklore and follows from Sealfon's approach \cite{Sealfon_2016} combined with the properties of zCDP shown by Bun and Steinke \cite{bun_steinke_2016}.

\begin{restatable}[\cite{Sealfon_2016} MST Error with Laplace Mechanism]{corollary}{lapM}
For any $\epsilon > 0$, probability $1 - \gamma$, and weights $\vec{W}$ for a graph $G$, the graph-based Laplace mechanism is $\epsilon$-differentially private and allows to compute an MST with approximation error at most $\mathcal{O}\left(\frac{4n}{\epsilon}\log (n/\gamma)\right)$ for the $\ell_1$-neighboring relationship and $\mathcal{O}\left(\frac{4n^{3}}{\epsilon}\log(n/\gamma)\right)$ for $\ell_\infty$.
\end{restatable}

\begin{restatable}[{\cite[Similar]{Sealfon_2016}} MST Error for Gaussian Mechanism]{corollary}{gaussM}
For any $\rho > 0$, probability $1 - \gamma$, and weights $\vec{W}$ for a graph $G$, releasing $\vec{W} + \normalExpD(0, \Delta_2^2/(2\rho))^{n \times n}$ is $\rho$-zero-concentrated dp and allows to compute an MST with approximation error at most $\errorGaussianNoiseloneExact$ in the $\ell_1$-neighboring relationship and $\errorGaussianNoiselinftyExact$ in $\ell_\infty$.
\end{restatable}

Contrary to post-processing, \textbf{in-place} algorithms inject the noise adaptively inside a concrete algorithm. 
It was shown that one can replace each step in Prim-Jarník's \cite{Pinot_2018} or Kruskal's \cite{McKenna_Miklau_Sheldon_2021} algorithm with a \textit{differentially private selection} mechanism like Report-Noisy-Max \cite{dwork_roth_2014}, the exponential mechanism \cite{Pinot_2018,Dwork_Nissim_Smith_2006}, or Permute-and-Flip \cite{mckenna_sheldon_2020}.
In this work, we will consider the approach proposed in Pinot's master thesis \cite{pinot_2018_ma}, where he introduced the \textit{Private Approximated Minimum Spanning Tree} (PAMST) algorithm.

\section{Faster Private Minimum Spanning Trees}

\subsection{Efficient Report-Noisy-Max}
As discussed in the technical overview, we use two shortcuts to simulate {\tt RNM} efficiently.
Both of them rely on the fact that discretizing weights down to some integer multiple of $s$, where $s>0$ is a parameter, is still $\epsilon$-DP if one adds noise drawn uniformly from ${\expD\left(\frac{\epsilon}{2(\Delta_\infty+s)}\right)}$.
Intuitively, rounding down increases the sensitivity by at most $s$, and therefore we need to scale by $(\Delta_\infty + s)$ instead of the usual scaling factor of $\Delta_\infty$ (see \cite{dwork_roth_2014}).
For our theoretical results, we will set $s = \Delta_\infty$ which does not decrease the utility much and allows us to use some computational shortcuts to speed up the sampling procedure.
The first step in our algorithm simulates {\tt RNM} on $m$ distinct edge weights with only $c \cdot \sqrt{n} \cdot \log n$ samples (if we pay for the $\mathcal{O}(m)$ initialization costs) with high probability.

\begin{restatable}[{\tt Discretized-RNM}]{corollary}{discRNM}\label{cor:privacy}
 For a graph $G = (V,E, \vec{W})$ with sensitivity $\Delta_\infty$ and for a quality (weight) function $w_G:E \rightarrow \R$, $s >0$ on $G$ with sensitivity $\Delta_\infty$, it satisfies $\epsilon$-DP, if we release
\[
    \arg\max_{e\in E}\left(\round{s}{w_G(e)} +\expD\left(\dfrac{\epsilon}{2(\Delta_\infty+s)}\right)\right)\enspace.
\]
\end{restatable}
\begin{proof} 
See \cref{apx:fast-pamst}.
\end{proof}

\subsection{Sampling Noise for (Top) Groups}\label{ch:top}
Assume we discretize all weights down to the next multiple of $\Delta_\infty$ and we already have a partition $F$ of the edges into groups $F_i \subseteq E$ where each pair $e,e'\in F_i$ has the same discretized weights.
The only edge within a group that can win {\tt RNM} is the edge with the maximum noise.
Because we draw independent exponential noise with the same scaling for each edge, we can alternatively only sample the maximum amount of noise $Z_i$ for each group $F_i$ and add $Z_i$ to a uniformly drawn element from $F_i$.
Then, we can release the overall $\argmax$.

We show that grouping without discretization matches the standard {\tt RNM} output distribution with exponential noise. 
More formally:
\begin{corollary}[Report-Noisy-Grouped-Max]\label{cor:equaldist}
Given a graph $G$ with edge sensitivity $\Delta_\infty$, privacy parameter $\lambda$, a weight function $w_G:E \rightarrow \R$ and partition of the edges $F := F_1 \uplus F_2 \uplus \cdots = E$ where for all $i$ and $\forall e,e'\in F_i: w_G(e) = w_G(e')$, and let $w_{F_i}$ denote the weight of edges in group $i$, then releasing an edge from
\[
\Uni \left(\argmax_{\substack{F_i \in F}}\left( w_{F_i} + \maxExpD(|F_i|, \lambda) \right)\right)
\]
has the same output distribution as {\tt Report-Noisy-Max} \cite{dwork_roth_2014}, where $\argmax _{e\in E}\left( w(e) + \expD(\lambda)\right)$ is being returned.
\end{corollary}

\begin{proof}

The standard {\tt RNM} returns $e'$ where $w(e') + \expD(\lambda)$ is the maximum among all noisy scores.
By the associativity of the max operator, the noise of $e'$ is also a maximum among all other possible outcomes $e^*\in E$ that have the same score $w(e') = w(e^*)$.
Hence, let $Y = \max(Z_1,\cdots, Z_{|F_i|})$, where $Z_i\sim \expD(\lambda)$, and the maximum within a group is distributed according to $Y$.
Then, by \cref{def:maxexpd}, we know that $Y$ follows exactly $\maxExpD(|F_i|, \lambda)$.
Since the noisy $\argmax$ inside a single $F_i$ is uniformly distributed, we can sample $Y_i \sim \maxExpD(|F_i|, \lambda)$ for each group $F_i$ and then add it to some uniformly chosen element in $F_i$.
\end{proof}

To benefit from this new way of sampling {\tt RNM}, we have to combine it with the discretization (\cref{cor:privacy}).
The idea is that if we discretize to an integer multiple of $s = \Delta_\infty$ we only have to sample a single noise term from $\maxExpD$ for a small number of top groups and can combine it with another sampling technique that captures the event that something far from the maximum gets a large amount of noise.

\subsection{Sampling Bottom Values}\label{ch:bottom}
Let ${\tilde{M} = \max_{e \in E'} w_G(e)}$ be the maximum weight in some subset (of cut edges) $E' \subseteq E$ of a graph~$G$.
We show an alternative way to sample the noise for $e \in E'$ where $w_G(e) \leq \tilde{M} - M$ for some $M \in \R^{+}$.
Denote the set of these edges by ${L = \{e\in E'\;|\; w_G(e) \leq \tilde{M} - M \}}$.

Observe that for any $e\in L$ to become the noisy maximum, the noise term \emph{must} exceed $M$.
Therefore, instead of adding noise to each of the edges, we can alternatively sample the number of noise terms $k$ that exceed $M$ first from $k \sim \binomialD(|L|, e^{-\lambda \cdot M})$ and then sample independently $Z_1,...,Z_k \sim \expD(\lambda)_{|\geq M}$ conditionally on being larger than $M$. 
The idea of efficiently sampling a set of random variables whose values exceed a threshold $M$ has previously been used in the context of sparse histograms~\cite{cormode_differentially_2012}, but our use of the technique is different since we are also interested in the value of the maximum noise.
By the memorylessness property of the exponential distribution, drawing  $Z \sim \expD(\lambda)_{|\geq M}$ is equivalent to drawing $Z \sim  (M + \expD(\lambda))$.
Finally, we add the $Z_i$'s to the weights of a uniformly random chosen subset of edges from $L$.
To formally prove that it works, we can (theoretically) clip the noise of all weights in $L$ to $M$ and then show that the joint probability densities of our approach exactly match with the adding $\max(M, \expD(\lambda))$ noise to each element.
Observe that all elements in $L$ are more than $M$ away from the maximum weight $\tilde{M}$, and in the standard {\tt RNM}, we would also add noise to the maximum edge. 
Therefore, clipping would not have any impact.
Later, we will show how to choose $M$ so that $k$ is very small with high probability.

\begin{restatable}{lemma}{bottom}\label{alg:alternative-bottom}
Given a graph $G = (V, E)$ with weights $w:E\rightarrow \R$, a privacy parameter $\lambda > 0$ and a threshold $M > 0$. 
Define $\tilde{M} := \max_{e\in E}(w(e))$ to be the maximum weight in $G$.
We consider a subset $L := \{e\in E| w(e) \leq \tilde{M} - M\}$.
Then, the vectors $\Vec{b} = (b_e)_{e\in L}$ and $\vec{b'} = (b_e')_{e\in L}$ are identically distributed if they are sampled from the following two mechanisms.

      \begin{itemize}
         \item \textbf{Clipped-RNM}: Release $b_e := w_G(e) + \max(M, \expD(\lambda))$ \label{alg:rnm-bottom}
         \item \textbf{Alternative-RNM}: \begin{enumerate}
            \itemsep-0.6em 
            \item First sample $k\sim \binomialD\left(L, \exp(-\lambda M)\right)$, 
            \item then draw uniformly a subset of edges $I \sim \UniformD\left(\binom{[L]}{k}\right)$ of size $k$, 
            \item and finally release $b_e' := w(e) + \begin{cases}
         M $\textbf{~if~}$ e \notin I {\color{gray}{\text{~\# clip noise to M}}}\\ 
         M + \expD(\lambda) $\textbf{~else~}$
         \end{cases}$
      \end{enumerate}\label{alg:rnm-bottom-fast}
   \end{itemize}
\end{restatable}

\begin{proof}[Proof]
See \cref{proof:bottom}.
\end{proof}

\subsection{Simulating Report-Noisy-Max: The Full Algorithm}

\begin{figure}
   \centering
    \includegraphics[width=0.5\textwidth]{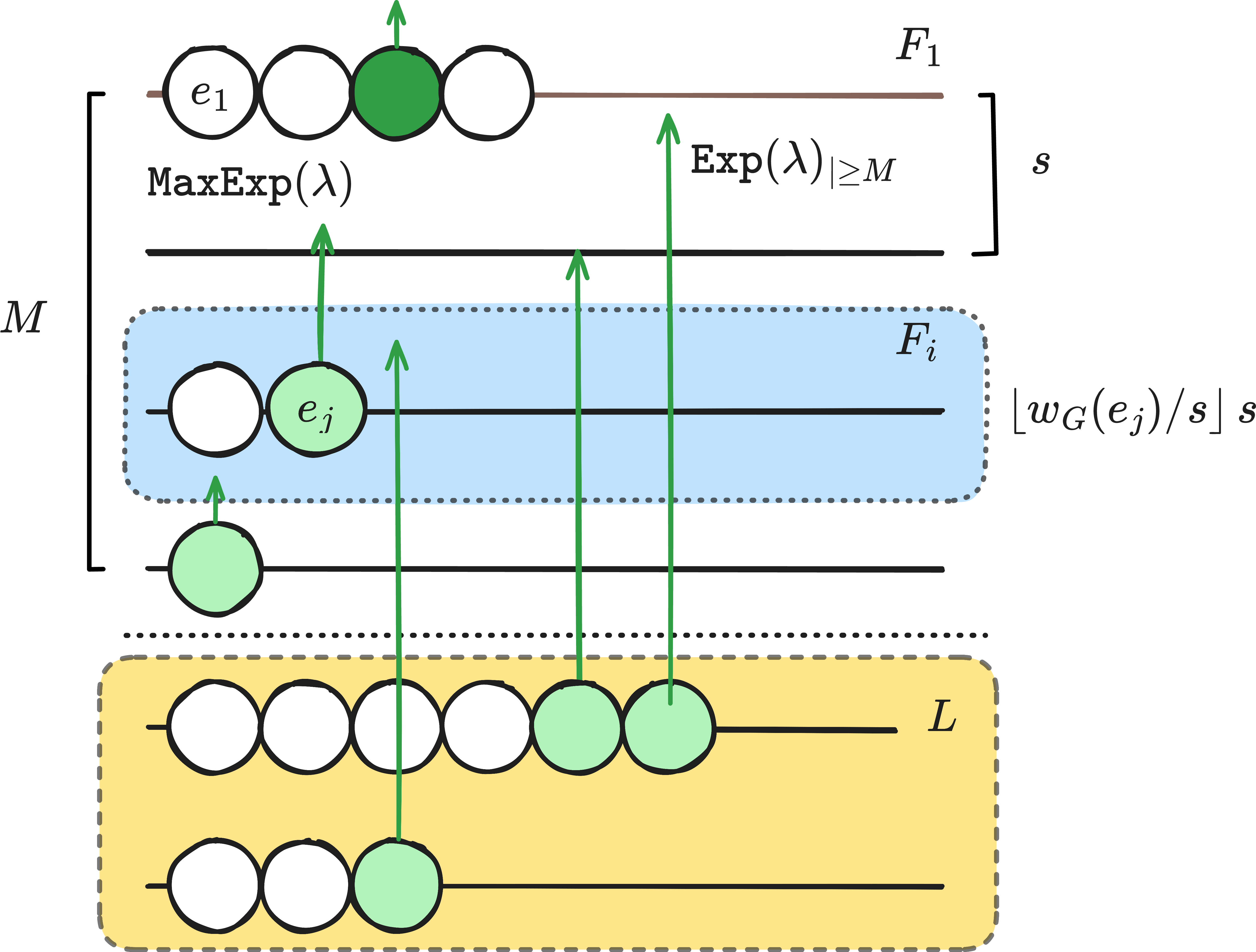} 
    \caption{
   Our fast {\tt RNM} simulation. 
   Balls on a horizontal line have been discretized to the same value and form a group $F_i$.
   For every edge below the threshold $M$ away from the maximum (denoted as the set $L$), we first sample the number of noise $z$ terms exceeding $M$ from $\binomialD(|L|, e^{-\lambda M})$, then conditionally sample $k$ noise terms from ${\tt Exp}(\lambda)_{|\geq M}$ and add them to the edges of random subset of size $k$
   We uniformly select an edge for each top group $F_i$ and add noise drawn from $\maxExpD(|F_i|, \lambda)$.
   The \ulcolor[lightgreen]{light green} elements are the only ones we have to sample noise for. 
   This example returns the \ulcolor[darkgreen]{dark green} edge.
   }
\label{fig:ouralg}
\end{figure}

We can now combine both sampling procedures to give the complete algorithm.
The idea is to discretize the edges into the groups $F_i$ as described in \cref{cor:privacy}, which we later do only once during initialization.
Then we split them to have at most $M = \max(1, \log(n) \sqrt{n} \Delta_\infty /\sqrt{\rho})$ top groups and the rest of the edges in $L$ as shown in \cref{alg:rnm-complete}. 
The running time depends on the underlying data structure and how $E'$ and the (discretized) weights are stored.
You can find a visualization of the complete algorithm in \cref{fig:ouralg}.

\begin{algorithm}[htb]
   \caption{\rnmCompleteFast}
   \label{alg:rnm-complete}
   \begin{algorithmic}[1]
      \parameters{Privacy parameter $\lambda$}
      \Require {Graph $G=(V,E)$, weights $w: E\rightarrow \R$, sensitivity $\Delta_\infty$, subset of (cut) edges $E' \subseteq E$}
      \State Discretize all $e\in E$ to $\tilde{w}(e) :=\roundInline{\Delta_\infty}{w(e)}$ and group them into $F_i$
      \State $M = 4 \sqrt{n} \log {n} \Delta_\infty/\sqrt{\rho}$ and $\tilde{M} = \max\limits_{e\in E'}\tilde{w}(e)$.
      \State Split edges into $L := \{e\in E'|\tilde{w}(e) <  \tilde{M} - M\}$
      \State Simulate noisy grouped max for groups in $E' \setminus L$ (as in  \cref{ch:top})
      \State Sample noise for bottom edges (as in \cref{alg:alternative-bottom})
      \State Return the overall $\argmax$
   \end{algorithmic}
\end{algorithm}

\begin{corollary}
   \cref{alg:rnm-complete} with $\lambda = \dfrac{\sqrt{2\rho}}{4\Delta_\infty}$ is $\rho$-zCDP.
\end{corollary}
\begin{proof}
   As for the top groups, we match the output distribution of {\tt Report-Noisy-Grouped-Max} (\cref{cor:equaldist}), and $L$ also simulates adding exponential noise exactly, the overall mechanism matches {\tt Discretized-RNM} (\cref{cor:privacy}).
   Therefore the whole mechanism is $\rho$-zCDP by applying the conversion theorem (\cref{lem:composition-zcdf}), and setting $s = \Delta_\infty$.
\end{proof}

\section{Efficiently Finding an MST}

We start by presenting a special priority queue that supports all the operations required by our MST  algorithm.

\begin{figure}[t]
   \centering
    \includegraphics[width=0.9\textwidth]{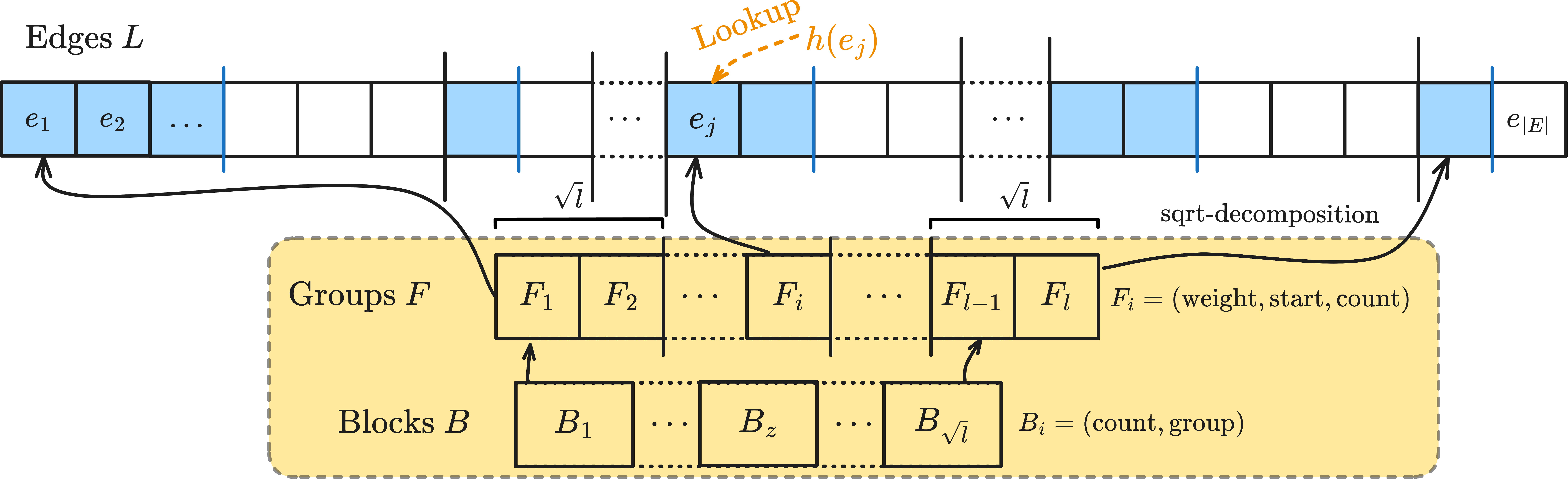} 
\caption{
   Visualization of our data structure for storing the discretized edges.
   $L$ contains all edges sorted by their (discretized) weight.
   All $F_i$'s point to an interval in $L$, where all its edges are stored. 
   Inserted edges are stored from the left inside each interval, and if a new one gets inserted or deleted, we swap them locally and update the corresponding counters.
   To find them in constant time, we need an additional dictionary $h:E\rightarrow \mathbb{N}$, which stores the current index in $L$ for each particular edge.
   Using a \emph{sqrt-decomposition} \cite{cpalgorithmsSqrtDecomposition}, we bundle $\sqrt{l}$ many groups and enable fast search by adding $\sqrt{l}$ many blocks. 
   This allows to find the maximum in $\mathcal{O}(\sqrt{l})$ and to get the tighter Fast-PAMST running time of $\mathcal{O}(m + \log n \sqrt{n}/\sqrt{\rho})$ \cref{thm:main}, we need a more general data structure with four layers each holding $\sqrt[4]{l}$ elements.
   }
   \label{fig:ourpq}
\end{figure}

\paragraph{A Special Priority Queue}\label{ch:specialPQ}
We will describe a special data structure with constant-time {\tt insertion} and {\tt deletion} directly supporting the previously introduced sampling procedure.
The data structure consists of a sorted list $L$ of all possible $m = O(n^2)$ edges $E$ and a second array that keeps track of all the groups $F_i$.
Each $F_i$ points to an interval in this list where edges with the same weight (groups $F_i$) are stored. 
We swap the edges locally inside these intervals for {\tt insertion} and {\tt deletion}. 
To enable fast access, we need to keep an additional dictionary that tracks the exact position of the edges inside the list.
Because updating the maximum on any deletion operation is non-trivial, only adding a single pointer is not enough.
To find the real maximum from which we start sampling for the following $M/\Delta_\infty$ groups, we use a technique similar to what is known as \emph{sqrt-decomposition}~\cite{cpalgorithmsSqrtDecomposition}, but extend it to four layers.
This technique requires at most $4\, m^{1/4}$ comparisons to find the maximum active edge and these pointers can be updated in constant time.
Assuming random access to a sorted list of edges, we initialize the data structure by a single traversal over all edges; therefore, initialization takes time $\mathcal{O}(m)$.
{\tt Insertion} and {\tt deletion} both take $\mathcal{O}(1)$, and we can directly use the previously introduced sampling mechanism on top of it, which runs in $\mathcal{O}(\sqrt{n} \log (n)/\sqrt{\rho})$ with high probability.
Prim-Jarník's algorithm on a dense graph has at most $n$ inserts and deletions per step and reports a noisy maximum once.
If we keep the insert and delete operations in $\mathcal{O}(1)$ time, we can allow linear time for a single execution of {\tt RNM}. 
Figure \ref{fig:ourpq} shows a data structure visualization.
We are now ready to prove \cref{thm:main}, restated here for convenience.

\main*

\begin{proof}
The idea is to replace the priority queue used in Prim-Jarník's algorithm \cite{Prim1957} with the data structure described above.
We must negate the weights to return a minimum spanning tree instead of a maximum.
We analyze \emph{privacy}, \emph{utility}, and \emph{running time} separately.

~~\textbf{Privacy.} 
 Prim-Jarník's algorithm consists of $n-1$ steps, and we can use composition to achieve overall $\rho$-zCDP.
If we set $\lambda = \frac{\sqrt{2\rho}}{4\sqrt{n-1}\Delta_\infty}$ then selecting each new edge using report noisy max is $\rho/(n-1)$-zCDP and the whole algorithm is $\rho$-zCDP.
As our algorithm simulates {\tt Report-Noisy-Grouped-Max} on the discretized graph exactly (see \cref{cor:equaldist}), the privacy argument from \cref{cor:privacy} immediately transfers.
By \cref{cor:privacy}, setting $\lambda = \frac{\epsilon}{4\Delta_\infty}$, is $\epsilon$-DP.
For $\rho > 0$ we can set $\lambda = \frac{\sqrt{2\rho}}{\sqrt{n}4 \Delta_\infty}$, we get $\rho/n$-zCDP for each call to {\tt RNM} and hence the overall algorithm is $\rho$-zCDP by composition.

~~\textbf{Utility.} To get the utility bound, we closely follow the argument of Pinot \cite{Pinot_2018}. 
For simplicity we assume $n>3$.
Denote by $w_T$ the MST weight of the output of \cref{alg:rnm-complete} and by $w_T^{*}$  the weight of the optimal MST.
By a union bound on $n-1$ steps and the utility of the {\tt RNM} (\cref{rnm:utility}) we have that, with probability $1- \mu$,
\begin{align*}
w_T - w_T^{*}  &\leq 4\cdot (n-1) \cdot \Delta_\infty\sqrt{n/(2\rho)}\cdot\log\left(n^{2}/\mu\right)\\
 &\leq n^{\nicefrac{3}{2}}\Delta_\infty\sqrt{2/\rho}\cdot\log\left(n^{2}/\mu\right)\enspace.
\end{align*}
Hence, we get the desired bound of $\errorPrivatePrimS$ for any $\mu$ that is polynomial in $1/n$.

~~\textbf{Running Time.} Prim has at most $n-1$ steps in each of which we add and remove at most $n$ edges from the set of cut edges.
In each step, we invoke our simulated version of {\tt RNM} once, which runs in $\mathcal{O}(\sqrt{n/\rho}\log(n))$ because finding the actual maximum takes $\mathcal{O}(\sqrt[4]{m})$ time (by using the sqrt-decomposition), we need a single sample for each of the  $M/\Delta_\infty$ groups, and sampling anything from the bottom $L$ happens with a tiny probability:
Trivially $|L|\leq m = O(n^{2})$, and by our choice of $M$, we can see that the probability that any edge in $L$ gets a lot of noise is minuscule.
Note that our algorithm is linear in the number of edges for sufficiently large $m$.  

If $M = c\log(n)/\lambda = 4c\log(n)\sqrt{n}\Delta_\infty/\sqrt{2\rho}$, for some suitable real $c > 0$, then by a union bound and the tails of the exponential distribution:
\begin{align*}
\Pr\limits_{X_1,...,X_{|L|} \sim \expD(\lambda)} \left[\exists X_i: X_i \geq \dfrac{c \log(n)}{\lambda}\right] &\leq n^{2}\cdot \exp\left(-\dfrac{c\log(n)\lambda}{\lambda}\right)
= n^{2-c}\enspace.
\end{align*}
Because of the discretization to $\Delta_\infty$, we get at most $c\sqrt{\frac{n}{2\rho}}\log(n)$ top groups $F_i$, for which we sample noise from $\maxExpD(|F_i|, \lambda)$ with probability $\frac{1}{n^{2-c}}$.
As initializing our data structure takes linear time in the number of edges and each edge is inserted and deleted into the priority queue at most once,  the total running time is $\mathcal{O}(m + n^{3/2} \log(n) /\sqrt{\rho})$. \qedhere
\end{proof}

\section{Empirical Evaluation}
 
To support our claims, we have implemented {\tt Fast-PAMST}, Pinot's
{\tt PAMST} \cite{Pinot_2018}, and Sealfon's Post-Processing \cite{Sealfon_2016} algorithms in C++.
Due to insufficient documentation of the codebase accompanying existing works \cite{Pinot_2018, pinot_2018_ma}, we implemented everything from scratch.
It was compiled with clang (version 15.0.0) and relies on the \emph{Boost} library (version 1.84.0).
Graphs are stored using the \emph{adjacency\_list} representation provided by \emph{Boost.Graph}.

The post-processing algorithm adds noise from $\normalExpD(0,\frac{n^2\Delta_\infty}{2\rho})$ to each $w_e$, then finds an MST on the noisy graph and calculates the error of these edges on the real weights.
To find a noisy maximum edge running {\tt PAMST}, we iterate over all active edges stored in a dictionary, add noise taken from $\expD(\frac{\sqrt{2\rho}}{2\sqrt{n}\Delta_\infty})$ to each weight and report the $\argmax$.
The {\tt Fast-PAMST} implementation follows closely the description in \cref{alg:Fast-PAMST}.
All experiments were run locally on a MacBook Pro with an \emph{Apple M2 Pro} processor (10 Cores, up to 3.7GHz) and 16GB of RAM. 
We ran each parameterization five times and plotted the median.

\paragraph{Results}

\begin{figure}
    \resizebox{\textwidth}{!}{\input{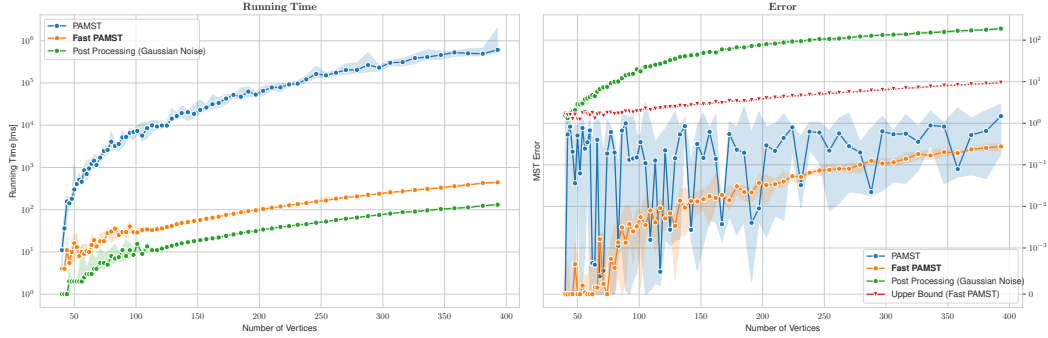}}
    \caption{Experiments on a complete graph with $n$ vertices for a a fixed privacy parameter $\rho = 0.1$ where each $w_e \sim \UniformD(0,1)$. Each data point is computed from a median of five runs (spanning the area around the curve) and runs on a MacBook Pro (16GB RAM, M2 Pro). }
    \label{fig:experiment1}
\end{figure}

The left plot of~\cref{fig:experiment1} shows the running time of the three algorithms on a complete graph $G = (V,E)$, where for each edge $e\in E$, we independently draw $w_e$ from $\UniformD(0,1)$, set $\rho = 0.1$ and $\Delta_\infty=10^{-5}$.
The privacy parameter $\rho$ is chosen to achieve a reasonable level of privacy, and $\Delta_\infty$ is low enough to get a good error while having enough distinct groups the algorithm must to consider.
One could encounter this type of data if by considering the mutual information matrix of an underlying dataset with approximately $1.4\cdot 10^6$ rows.

As expected, our algorithm closely follows the asymptotic running time of the post-processing approach, outperforming the {\tt PAMST} algorithm.
The slight upward shift of {\tt Fast-PAMST} might be due to the additional effort of initializing the priority queue and constant overhead in each of the more complex data structure modifications.
The right plot shows the absolute error of the cost of the found MST to the optimal MST.
One can see that the error for both {\tt PAMST} and {\tt Fast-PAMST} is very small.
Observe that the {\tt Post-Processing} algorithm approaches roughly $n/2$ due to the high amount of noise required.
As the amount of noise scales with $n^2$, we quickly release a nearly random tree.
The error for both {\tt PAMST} and {\tt Fast-PAMST} is extremely low, and {\tt Fast-PAMST} has a noticeably lower error for smaller graphs.
It seems that {\tt Fast-PAMST} behaves better than expected for smaller graphs, but for larger graphs, {\tt Fast-PAMST} and {\tt PAMST} resemble each other, which is an indication that they exhibit the same asymptotic error behavior.
Furthermore, we added the high probability upper bound for {\tt Fast-PAMST} (See \cref{thm:main}), and it seems that this bound is very loose for this setting.

We also ran {\tt Fast-PAMST} on larger graph instances, which can be found in \cref{fig:experiment2}.

\newpage
\section{Conclusion and Open Problems}
We have seen that, at least for sufficiently dense graphs, it is possible to privately compute an approximate MST (under the $\ell_\infty$ neighborhood relation) such that we simultaneously achieve: asymptotically optimal error \emph{and} running time linear in the size of the input.
It would be interesting to extend this ``best of both worlds'' result to sparse graphs, as well as to general $\ell_p$ neighborhood relations.

Following the approach by McKenna, Miklau, and Sheldon \cite{McKenna_Miklau_Sheldon_2021}, we could further investigate the application of Chow-Liu \cite{chow_liu_1068} trees where the bottleneck is the computation of the mutual information matrix.
Our efficient simulation technique for {\tt RNM} is fairly general, and it would also be nice to find applications in other contexts.

\newpage
\bibliographystyle{plain}
\typeout{} 
\bibliography{references}

\newpage
\appendix

\onecolumn
\section{Appendix}

\subsection{Preliminaries}\label{apx:defs}

\paragraph{Graph Theory.} A~{\textit{spanning tree}} is an acyclic subset $E'\subseteq \E$ where $V = \bigcup\limits_{e' \in E'} e'$.
A \textit{minimum (cost) spanning tree} (MST) is a spanning tree $T^*$ where $W_{T^*}$ is minimum among all other spanning trees $T'\subseteq E$.
Throughout the work, assume that $G$ has at least two different spanning trees.

\paragraph{Differential Privacy.}

Intuitively, an algorithm is \textit{private} if slight changes in the input do not significantly change the probability of seeing any particular output.
\begin{definition}[\cite{Dwork_Nissim_Smith_2006} $(\epsilon, \delta)$-differential privacy]\label{def:priv}
Given $\epsilon > 0$ and $\delta \geq 0$, a randomized mechanism $\mathcal{M}$ satisfies ($\epsilon$, $\delta$)-DP if and only if for every pair of neighboring weight matrices $\vec{W} \sim \vec{W'}$ on a particular graph topology $G$, and for all possible outputs $Z$:
\[
    \Pr[\mathcal{M}(\vec{W}) = Z] \leq e^\varepsilon \Pr[\mathcal{M}(\vec{W'}) = Z] + \delta\enspace.
\]

If $\delta = 0$, $M$ satisfies \textit{pure differential privacy} and \textit{approximate differential privacy} otherwise.
\end{definition}

This work uses $\rho$-\textit{zero-Concentrated Differential Privacy} introduced by Bun and Steinke \cite{bun_steinke_2016}.

\begin{definition}[\cite{bun_steinke_2016} $\rho$-zero-Concentrated Differential Privacy (zCDP)]
Let $\mathcal{M}$ denote a randomized mechanism satisfying $\rho$-zCDP for any $\rho > 0$. 
Then for all $\alpha > 1$ and all pairs of neighboring datasets $\vec{W} \sim \vec{W'}$, we have
\[ 
    D_\alpha\left(\mathcal{M}(\vec{W}) \vert \vert \mathcal{M}(\vec{W'})\right) \leq \rho \alpha, 
\]
where $D_\alpha(X \vert \vert Y)$ denotes the $\alpha$-Rényi divergence between two distributions $X$ and $Y$.
\end{definition}

\begin{lemma}[\cite{bun_steinke_2016} Composition and Conversion]\label{lem:composition-zcdf}
    If $M_1$ and $M_2$ satisfy $\rho_1$-zCDP and $\rho_2$-zCDP, respectively.
    Then $M=(M_1, M_2)$ satisfies $(\rho_1 + \rho_2)$-zCDP.
    If $\mathcal{M}$ satisfies $\rho$-zCDP, then $\mathcal{M}$ is $(\varepsilon, \delta)$-DP for any $\delta > 0$ and $\varepsilon=\rho + 2\sqrt{\rho \log(1/\delta)}$.
\end{lemma}

For {\tt Report-Noisy-Max}, we also need the notion of a utility function's $p$-sensitivity $\Delta_p$.
\begin{definition}[$p$-sensitivity of a function $\Delta_p$]\label{psensitivefunction}
        For a utility function  $w: E' \times \R^{n \times n} \rightarrow \mathbb{R}^{d}$ on a set $E'\subseteq E$ and $d\in \mathbb{N}$ and neighboring $\vec{W}\sim \vec{W'}$, we denote the $\Delta_p$-sensitivity of $w$ as
    \[
        \Delta_p(w) :=\max_{e\in E'} \max_{\vec{W} \sim \vec{W'}} \vert\vert w(e,\vec{W}) - w(e', \vec{W'})\vert\vert_p \enspace .
    \]
\end{definition}

\subsection{Analysis of Post-Processing MST algorithms}\label{ch:analysis}

\lapM*
\begin{proof}
See Sealfon's proof \cite{Sealfon_2016} for $\ell_1$, which is a combination of a union bound on the $n-1$ selected edges of the spanning tree and a concentration bound on the noise $\lapD(\Delta_1/\epsilon)$ for each edge.
For $\ell_\infty$, we have to scale the noise by $X_e \sim \lapD(n^{2}/\epsilon)$ for each edge which concentrates around $\vert X_e\vert \leq (2n^{2}/\epsilon)\log(n/\gamma)$ with probability $1-\gamma$.
Reusing Sealfon's $\ell_1$ arguments again, we get an error of $\mathcal{O}(4n^{3}/\epsilon \log(n/\gamma))$.
The overall mechanism is $\epsilon$-differentially private as the Laplace mechanism is $\epsilon$-dp and the computation of the MST is just post-processing.
\end{proof}

Instead of using Laplace noise, we can switch to the Gaussian mechanism.

\gaussM*
\begin{proof}
\textbf{Privacy} for both $\ell_1$ and $\ell_\infty$-neighboring relationships follows immediately from the Gaussian mechanism \cite{Dwork_Nissim_Smith_2006} and the \textit{post-processing} property under $\rho$-zCDP \cite{bun_steinke_2016}. 
For proving the \textbf{utility} bound, let $\vec{W'} = \vec{W} + \vec{Z}$ where $\vec{Z}\sim \normalExpD(0, \Delta_2^2/2\rho)^{n \times n}$ be the private weight matrix.
Combining a union with the Chernoff bound, gives $\vert Z_{i,j}\vert \leq \Delta_2 \sqrt{\log(n^2/\gamma)/\rho}$ w.p $1 - \gamma$ for all $i,j \in [n]$.
Note that $\Delta_2 \leq \sqrt{n^{2}\Delta_\infty} = n\sqrt{\Delta_\infty}$ under the $\ell_\infty$ neighboring relationship and $\Delta_2 \leq \sqrt{\Delta_\infty}$ under $\ell_1$.

 Denote $W_T$ as the real cost if the edges of the MST are computed on $\vec{W}_\text{priv}$ and $W_{T^*}$ the cost of the optimal MST.
 We follow the same chain of inequalities as Sealfon used \cite{Sealfon_2016}, we get  
 \[
  W_T \leq W_{T^{*}} + 2n^2\sqrt{\Delta_\infty\log\left(n^2/\gamma\right)/\rho} 
 \]
 for $\ell_\infty$, and
 \[
  W_T \leq W_{T^{*}} + 2n\sqrt{\Delta_\infty\log\left(n^2/\gamma\right)/\rho} 
 \]
 for $\ell_1$, with probability probability $1-\gamma$.
    
\end{proof}
\subsection{Fast-PAMST}\label{apx:fast-pamst}

\subsubsection{Discretized Report-Noisy-Max}
\discRNM*
\begin{proof}
   Denote the {\tt Discretized-RNM}-mechanism proposed in \cref{cor:privacy} on a graph as $M$, and let $w:E \rightarrow \R$ be a $\Delta_\infty$-sensitive query and two neighboring graphs $G$ and $G'$ with  $\vec{W} \sim_\infty \vec{W'}$.
   For notational convenience, we shortly write $x_e := w_G(e)$, and $x_e' := w_{G'}(e)$ and analogously for the rounded values: $\tilde{x}_e := \roundInline{s}{w_G(e)}$ and $\tilde{x}_{e'} := \roundInline{s}{w_{G'}(e)}$ for each possible output $e \in E$ and $s > 0$.
   Let $\lambda = \frac{\epsilon}{2(\Delta_\infty+s)}$.
   For any single possible output $e \in E$, with fixed noise terms $Z_{e'} \sim \expD(\lambda)$ for all $e' \in E \setminus \{e\}$, we have:

   \begin{align}
   \Prob\left[M(G) = e\right]
   &= \Prob_{Z_e \sim \expD(\lambda)}\left[Z_e \geq \max_{e' \in E} \tilde{x}_{e'} - \tilde{x}_e + Z_{e'}\right] \nonumberinc \\
   &= \Prob_{Z_e \sim \expD(\lambda)}\left[Z_e \geq \max_{e' \in E} \left(\round{s}{x_{e'}} - \round{s}{x_e} + Z_{e'}\right)\right] \nonumberinc \\
   &\leq \Prob_{Z_e \sim \expD(\lambda)}\left[Z_e \geq \max_{e' \in E} \left(\left(x_{e'} - s\right) - x_e + Z_{e'}\right)\right] \label{line:round}\\
   &\leq \Prob_{Z_e \sim \expD(\lambda)}\left[Z_e \geq  \max_{e' \in E} \left(x_{e'}' - \left(\round{s}{x_e'} - s\right) + Z_{e'}\right) - 2 (\Delta_\infty + s)\right] \label{line:sensitivitybound} \\
   &= e^{2\lambda (\Delta_\infty + s)} \Prob_{Z_e \sim \expD(\lambda)}\left[Z_e \geq \max_{e \in E} \left(x_{e'}' -\round{s}{x_e}  + Z_{e'}\right)\right] \nonumberinc \\
   &\leq e^{\lambda 2(\Delta_\infty + s)} \Prob_{Z_e \sim \expD(\lambda)}\left[Z_e \geq \max_{e \in E} \left( \round{s}{x_{e'}'}\tilde{x}_{e'}' -\round{s}{x_e'}+ Z_{e'}\right)\right] \label{line:gettingback} \\
   &= e^{\lambda 2(\Delta_\infty + s)} \Prob\left[M(G') = e\right] \nonumberinc \enspace.
   \end{align}
   Hence, for $\lambda = \frac{\epsilon}{2(\Delta_\infty + s)}$, $M$ is $\epsilon$-DP.
\end{proof}

The proof uses the fact that the overall probability increases by decreasing the value inside the $\max$ and additionally uses some inequalities that the rounding induces and holds for all $e,e'\in E$
\Cref{line:round} holds, because $\roundInline{s}{x_{e'}} \leq x_{e'} - s$ and $-\roundInline{s}{x_e} \leq - x_e$.
\Cref{line:sensitivitybound} holds by applying the sensitivity $|x_e-x_{e}'|\leq \Delta_\infty$ twice, and hence $-x_e \geq -\Delta_\infty - x_e'$ and by symmetry also $x_{e'} - x_{e'}' \geq -\Delta_\infty \Leftrightarrow x_{e'} \geq -\Delta_\infty + x_{e'}'$.
It holds that $x_e \leq \roundInline{s}{x_e'} + s \Leftrightarrow -x_e' \geq -\roundInline{s}{x_e'} - s$.
Finally, \cref{line:gettingback} follows from  $\roundInline{s}{x_e}\leq x_e$.

\subsubsection{Alternative Sampling for Bottom}\label{proof:bottom}
We now formally prove of why our alternative sample procedure for values below some threshold works.
\bottom*
\begin{proof}
   We prove the claim by proving that the vectors $\vec{b}=(b_e)_{e\in L}$ and $\vec{b'}=(b_e')_{e\in L}$ are identically distributed by showing that their joint PDF is equal.
   Let $M > 0$ be a threshold, and ${L := \{e\in E|w(e) \leq \tilde{M} - M\}}$.
   For all $e \in E$, Let $b_e$ and $b_e'$ be random variables of {\tt Clipped-RNM} and {\tt Alternative-RNM} as described in \cref{alg:alternative-bottom} respectively.
   For all $e\in L$ let $\alpha_e \in \R$ and $Z_e \sim \expD(\lambda)$ .
   Furthermore, without loss of generality, we can assume that $w(e) + M \leq \alpha_e$. 
   Hence, we get for {\tt Clipped-RNM}:

   \begin{align*}
      \Prob\left[\bigwedge_{e \in L} b_e \geq \alpha_e\right] &= \prod_{e \in L} \Prob[b_e > \alpha_e] \\
      &=  \prod_{e \in L} \left(\Prob[Z_e < M \wedge \overbrace{w(e) + Z_e}^{b_e} \geq \alpha_e] + \Prob[Z_e\geq M \wedge \overbrace{w(e) + Z_e}^{b_e} \geq \alpha_e] \right) \\
      &= \prod_{e \in L} \Prob\left[w(e) + Z_e > \alpha_e | Z_e \geq M\right] \cdot \Prob\left[Z_e \geq M\right] \\
      &= \prod_{e \in L} \Prob[Z_e > \alpha_e - w(e) - M]\cdot \exp(-\lambda M) \\
      &= \exp\left(-\lambda \sum_{e \in L} (\alpha_e - w(e))\right) \enspace.
   \end{align*}
   The third line follows from the fact that $\Prob\left[Z_e < M \wedge b_e \geq \alpha_e\right] = 0$ by assumption.
   The last step follows from the memoryless property of the exponential distribution and its exact tails.
   Now, let $I$ be a uniformly drawn subset of edges in $L$ of size $k \sim \binomialD(|L|, -\lambda M)$.
   Then, we get for {\tt Alternative-RNM}:

   \begin{align*}
      \Prob\left[\bigwedge_{e \in L} b'_e \geq \alpha_e\right] &= \prod_{e \in L} \Prob[b'_e > \alpha_e] \\
      &= \sum_{I^* \subseteq L}\left(\Prob\left[I = I^*\right] \cdot \Prob\left[\bigwedge_{e \in L} b'_e > \alpha_e | I = I^*\right]\right) \\
      &= \Prob[I = L]\cdot \Prob\left[\bigwedge_{e \in L} b'_e > \alpha_e| I = L\right] \\
      &=\exp(-|L|\lambda M) \cdot \prod_{e \in L} \exp\left(-\lambda (\alpha_e - w(e) - M)\right) \\
      &=\prod_{e \in L} \exp\left(-\lambda (\alpha_e - w(e) - M)\right) \cdot \exp(-\lambda M) \\
      &= \exp\left(-\lambda \sum_{e \in L} (\alpha_e - w(e))\right) \enspace.
   \end{align*}

   The third line holds, because if $I^* \neq I$ then exists $e \in L$ such that  $b'_e = w(e) + M$ by definition and we know that $\Prob[\bigwedge_{e \in L} b_e' > a_e | I = I^*] = 0$ contradicting our assumption $w(e) + M \leq \alpha_e$.

   Furthermore, in line four, we use that $\Prob[I = L] = \Prob_{k \sim \binomialD(|L|, \exp(-\lambda M))}[k = |L|] = \exp(-|L|\lambda M)$,
   Hence $\Prob\left[\bigwedge_{e \in L} b_e \geq \alpha_e\right] = \Prob\left[\bigwedge_{e \in L} b'_e \geq \alpha_e\right]$ and $\vec{b}$,$\vec{b'}$ are identically distributed.
\end{proof}

\subsection{The Full {\tt Fast-PAMST} Algorithm}

\cref{alg:Fast-PAMST} shows our full variant of Prim-Jarník's algorithm where all the cut edges are stored in a special private priority queue.
Every time we pop a noisy max, we will iterate over all neighboring edges from the two endpoints, add those that have not been visited before into the priority queue, and immediately remove those that have been.
Finally, we will mark the start vertex to be visited.
Each step consists of at most $n$ inserts and deletes and exactly one pop\_noisy\_max operation.
Both sides are required to cover the initialization case.

    \begin{algorithm}
    \caption{\tt Fast-PAMST}
    \begin{algorithmic}[1]
    \parameters{privacy parameter $\rho$}
    \Require A graph \G with weights $w:E\rightarrow \R$ and sensitivity $\Delta_\infty$
    \State Initialize empty pq $P$ (as described in \cref{ch:specialPQ}) with discretization $\tilde{w}(e) := \roundInline{\Delta_\infty}{w(e)} $
    \State Let $v:=V_0$ for an arbitrary start vertex $V_0\in V$
    \State ${\tt visited} = \{v\}$, $S_E = \emptyset$
    \Statex ~
    \While {$|{\tt visited}| < |V|$}
    \State Add all edges $e$ adjacent to $v$ into $P$ for which  $e \setminus {\tt visited} = \{v\}$
    \State Remove all edges $e$ adjacent to $v$ to $P$ for which $e \setminus {\tt visited} = \emptyset$

    \State $e \leftarrow$ report noisy maximum edge from $P$ (using \cref{alg:rnm-complete})
    \State Remove $e$ from $P$
    \State $v := e \setminus {\tt visited}$

            \State ${\tt visited} \leftarrow {\tt visited} \cup \{v \}$
        \State $S_E \leftarrow S_E \cup \{ e \}$
    \EndWhile
    \Return $S_E$
    \end{algorithmic}
    \caption*{{\normalfont \bf Fast-PAMST: }
    }
    \label{alg:Fast-PAMST}
    \end{algorithm}

\newpage
\clearpage
\subsection{More Empirical Results}

\begin{figure}[h]
    \resizebox{\textwidth}{!}{\input{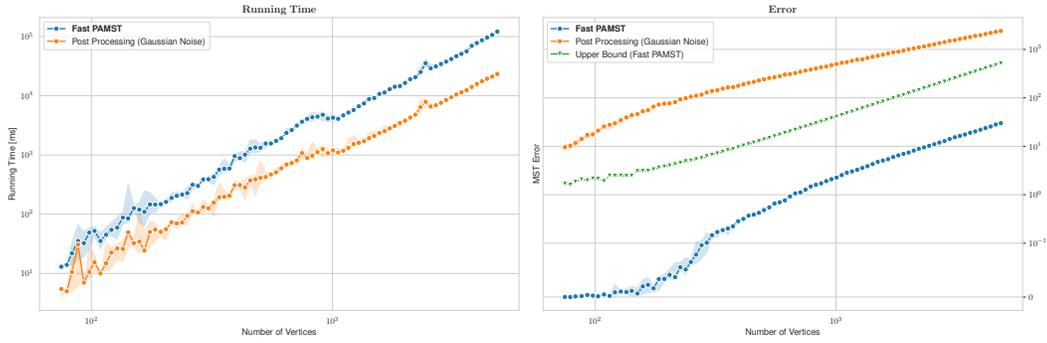}}
    \caption{Experiments on a complete graph with $30 \leq n \leq 5000$ vertices for a a fixed privacy parameter $\rho = 0.1$ where each $w_e \sim \UniformD(0,1)$. 
    Each data point is computed from a median of five runs (spanning the area around the curve) and ran on a MacBook Pro (16GB RAM, M2 Pro).
    }
    \label{fig:experiment2}
\end{figure}

\cref{fig:experiment2} shows an experiment for larger graphs.
We can see that  \emph{Post Processing} line quickly approaches $n/2$, which is exactly the expectation of any uniformly drawn spanning tree in $G$.
On the contrary, \emph{Fast-PAMST} on one side gets away with a very small error and, on the other side, follows the provided \emph{upper bound}.

\newpage
\clearpage

\end{document}